\documentclass[showkeys,aps,showpacs,amssymb,prd]{revtex4}

\usepackage{appendix}
\usepackage{amsbsy}
\usepackage{amsfonts}
\usepackage{amsmath}
\usepackage{amssymb}
\usepackage{amsthm}

\usepackage{graphicx}
\usepackage{ifthen}
\usepackage{textcomp}

\newcounter{multi} \newcounter{multa}

\newcounter{faki} \newcounter{faka}

    \newtheorem{theorem}{Theorem}
    
    \newtheorem{proposition}[theorem]{Proposition}
    \newtheorem{corollary}[theorem]{Corollary}

\theoremstyle{definition} 

\theoremstyle{remark} 

\def\bbf{{\bf f}}
\def\bc{{\bf c}}
\def\bx{{\bf x}}

\def\bs{{\bf s}}

\def\det {\mathop{\rm det}\nolimits}
\def\grad {\mathop{\rm grad}\nolimits}
\def\Jac {\mathop{\rm Jac}\nolimits}
\def\Hess {\mathop{\rm Hess}\nolimits}

\newcommand{\beqa}{\begin{eqnarray}}
\newcommand{\beq}{\begin{equation}}
\newcommand{\eeqa}{\end{eqnarray}}
\newcommand{\eeq}{\end{equation}}

\newcommand{\bo}{\boldsymbol}
\newcommand{\oli}{\overline}

\newcommand{\cideal}{\mathbb{C}[x]/(\varphi(x))}
\newcommand{\rideal}{R/(\varphi(x))}

\newcommand{\fkM}{\mathfrak{M}}
\newcommand{\fkm}{\mathfrak{m}}

\begin{document}

\title{A Universal Magnification Theorem II. Generic Caustics up to Codimension Five}

\author{A. B. Aazami}\email{aazami@math.duke.edu}\affiliation{Department of
Mathematics, Duke University, Science Drive, Durham, NC 27708}

\author{A. O. Petters}\email{petters@math.duke.edu}\affiliation{Departments of
Mathematics and Physics, Duke University, Science Drive, Durham, NC 27708}


\begin{abstract}
We prove a theorem about magnification relations for all generic general caustic singularities 
up to codimension five: folds, cusps, swallowtail, elliptic umbilic, hyperbolic umbilic, butterfly, parabolic umbilic, wigwam, symbolic umbilic, $2^{\rm{nd}}$ elliptic umbilic, and 
$2^{\rm{nd}}$ hyperbolic umbilic.  Specifically, we prove that for a generic
family of general mappings between planes exhibiting any of these singularities, and for a point in the target lying anywhere in the region giving rise to the maximum number of real pre-images
(lensed images), the total signed magnification of the pre-images will always sum to zero.  The
proof is algebraic in nature and makes repeated use of the Euler trace formula.
We also prove a general algebraic result
about polynomials, which we show yields an interesting corollary about Newton sums that in turn
readily implies the Euler trace formula. 
The wide field imaging surveys slated to be conducted by the Large Synoptic Survey Telescope
are expected to find observational evidence for many of these higher-order caustic singularities.
Finally, since the results of the paper are for generic general mappings, not just generic lensing maps, the findings are expected to be applicable not only 
to gravitational lensing, but to any system in which these singularities appear.

\end{abstract}

\pacs{98.62.Sb, 95.35.+d, 02.40.Xx}
\keywords{Gravitational lensing, caustics, substructure}

\maketitle
\section{Introduction}
\label{Introduction}
One of the key signatures of gravitational lensing is the occurrence of multiple images
of lensed sources. The magnifications of the images in turn are also known to obey certain relations.  These relations fall into two types: ``global'' and ``local''.  ``Global'' magnification relations involve all the images of a given source, but they are not {\it universal} because the relations depend on the specific class of lens models used.  Examples of such relations can be found in Petters et al. 2001 \cite[p. 191]{Petters}, Witt \& Mao 1995 \cite{Witt-Mao}, Rhie 1997 \cite{Rhie}, Dalal 1998 \cite{Dalal}, Witt \& Mao 2000 \cite{Witt-Mao2}, Dalal \& Rabin 2001 \cite{Dalal-Rabin}, and Hunter \& Evans 2001 \cite{Hunter-Evans}.  As shown in Werner 2007 \cite{Werner}, such relations are in fact topological invariants.

By contrast, ``local'' magnification relations are universal, but they apply only to a subset of the total number of images produced.  Two well-known examples of local magnification relations are the {\it fold} and {\it cusp} relations.  For a source near a fold or cusp caustic, the resulting images close to the critical curve are close doublets and triplets whose signed magnifications always sum to zero 
(e.g., Blandford \& Narayan 1986 \cite{Blan-Nar}, Schneider \& Weiss 1992 \cite{Sch-Weiss92},
Zakharov 1999 \cite{Zakharov}, \cite[Chap. 9]{Petters}).  The universality of these relations means that they hold independently of the choice of lens model.  In addition, 
the fold and cusp relations  have been shown to provide
powerful diagnostic tools for detecting dark substructure on galactic
scales using quadruple lensed images of quasars (e.g., Mao \& Schneider 1998  \cite{Mao-Sch}, Keeton, Gaudi \& Petters  
2003 and 2005 \cite{KGP-cusps,KGP-folds}). 

Recently, Aazami \& Petters 2009 \cite{Aazami-Petters}, which we consider Paper I,
established a universal magnification theorem for some of the higher-order caustics beyond folds and cusps, namely, the {\it swallowtail}, {\it elliptic umbilic}, and {\it hyperbolic umbilic} singularities.  These are generic caustic surfaces or big caustics occurring
in a three-parameter space.  Slices of the big caustics give rise to generic caustic metamorphoses
(e.g., \cite{Petters}, Chapters 7 and 9), all of which occur in gravitational lensing
(e.g., Blandford 1990 \cite{Blandford90}, Petters 1993 \cite{Petters93}, 
Schneider, Ehlers, \& Falco 1992 \cite{Sch-EF}, and \cite{Petters}).  It was shown in \cite{Aazami-Petters} that for lensing maps close to elliptic umbilic and hyperbolic umbilic caustics, and for general mappings exhibiting swallowtail, elliptic umbilic, and hyperbolic umbilic caustics, the total signed magnification for a source lying anywhere in the region giving rise to the maximum number of lensed images, is identically zero.  As an application, they used the hyperbolic umbilic to show how such magnification relations may be used
for substructure studies of four-image lens galaxies. 

The proof of these relations in \cite{Aazami-Petters} was elementary, but long, and thus was not amenable to higher-order caustics beyond the three mentioned above.  An elegant geometric technique, based on Lefschetz fixed point theory, has since been employed on these three singularities by Werner 2009 \cite{Werner2}.  The aim of our current paper is to extend these results to all the
remaining higher-order generic general singularities up to codimension $5$. In other words,
our findings are expected to be applicable not only to gravitational lensing, but to any system where these
singularities appear.

We prove that to {\it each} generic general caustic singularity
of codimension up to $5$---not just the fold, cusp, swallowtail, elliptic umbilic, and hyperbolic umbilic, but also the {\it butterfly, parabolic umbilic, wigwam, symbolic umbilic,} $2^{nd}$ {\it elliptic umbilic,} and 
$2^{nd}$ {\it hyperbolic umbilic}---is associated a magnification sum relation of the form
$$
\sum_{i} \fkM_{i} = 0\ .
$$
In other words, for generic families of general mappings between planes exhibiting these singularities, and for a point anywhere in the region of the target space giving rise to the maximum number of lensed images, the total signed magnification is identically zero.  Furthermore, as emphasized in \cite{Aazami-Petters}, magnification sum relations are in fact {\it geometric} invariants, because they are the reciprocals of Gaussian curvatures at critical points. 

Shin \& Evans 2007 \cite{Shin-Evans} constructed a realistic lens model for the Milky Way Galaxy and showed that it exhibited butterfly caustics (see also Evans \& Witt \cite{Evans-Witt} for another class of lens models that exhibit butterfly caustics).  More recently, Orban de Xivry \& Marshall 2009 \cite{Or-Marshall} created an atlas of predicted 
gravitational lensing due to galaxies and clusters of galaxies that can exhibit several of
these higher-order caustic singularities, and estimated the probabilities for their
occurrence.  They showed how a galaxy lens with 
a misaligned disk and bulge can generate swallowtails and butterflies, two merging galaxies
or galactic binaries can produce elliptic umbilics, and galactic clusters can create hyperbolic
umbilics.  These lensing effects are expected to be seen by the 
Large Synoptic Survey Telescope.

Concerning the tools of the paper, we mention that
the Euler trace formula was employed in \cite{Dalal-Rabin} to determine
 ``global'' magnification relations for {\it special} classes of lens models.
They used an analytical approach whereby they derived the Euler trace formula using residue calculus.
We show that the Euler trace
formula also lends itself quite naturally to ``local'' magnification relations for 
generic general
caustics, not just those occurring in gravitational lensing.
In fact, along with our main theorem, we also prove a general algebraic theorem about polynomials (Proposition~\ref{prop:recursive})
and derive a result about Newton sums as a corollary, which we show implies the Euler trace
formula.  Our main theorem is not a direct consequence of the Euler-Jacobi formula, of multi-dimensional residue integral methods, or of Lefschetz fixed point theory, because some of the singularities have fixed points at infinity.

The outline of the paper is as follows.  Section~\ref{Singularities} reviews the necessary singular-theoretic terminologies and results.  Section~\ref{Theorem} states our main theorem, which is for general mappings.  As preparation for the proof of our main theorem, in
Section~\ref{sec:recursion} we establish a recurrence relation for the coefficients of
the unique polynomials in cosets of certain quotient rings and show that this relation
yields a fact about Newton sums which can be employed to readily obtain the Euler trace formula.
We then use the results of Section~\ref{sec:recursion} to prove the main 
theorem in Section~\ref{Proof}.

\section{Higher-Order Caustic Singularities}
\label{Singularities}
In what follows, the term ``universal'' or ``generic'' is used to denote a property that holds for an open, dense subset of mappings in the given space of mappings.  With that said, consider a smooth, $n$-parameter family $F_{{\bo c},{\bo s}}({\bo {\rm x}})$ of functions on an open subset of $\mathbb{R}^2$ that induces a smooth $(n-2)$-parameter family of mappings $\bbf_{\bo c}({\bo {\rm x}})$ between planes ($n \geq 2)$.  Given $\bbf_{\bo c}({\bo {\rm x}}) = \bs$, call ${\bo {\rm x}}$ the {\it pre-image} of the target point $\bs$.  Critical points of $\bbf_{\bo c}$ are those ${\bo {\rm x}}$ for which ${\rm det}({\rm Jac}\,\bbf_{\bo c})({\bo {\rm x}}) = 0$.  Generically, the locus of critical points forms curves called {\it critical curves}.  The value $\bbf_{\bo c}({\bo {\rm x}})$ of a critical point ${\bo {\rm x}}$ under $\bbf_{\bo c}$ is called a {\it caustic point}.  These typically form curves, but could be isolated points.  Varying ${\bo c}$ causes the caustic curves to evolve with ${\bo c}$.  This traces out a caustic surface, called a {\it big caustic}, in the $n$-dimensional space $\{{\bo c},{\bo {\rm s}}\} = \mathbb{R}^{n-2} \times \mathbb{R}^2$.  Beyond the familiar folds and cusps, these surfaces form higher-order caustics that are classified into {\it universal} or {\it generic} types for locally stable families $\bbf_{\bo c}$.  Generic ${\bo c}$-slices of these big caustics are commonly called {\it caustic metamorphoses}.  

The {\it universal} form of the $(n-2)$-parameter family $\bbf_{\bo c}$ is obtained by using $F_{{\bo c},{\bo s}}$ to construct catastrophe manifolds that are projected into the space $\{{\bo c},{\bo {\rm s}}\} = \mathbb{R}^{n-2} \times \mathbb{R}^2$ to obtain local coordinates for $\bbf_{\bo c}$ (e.g.,  Majthay 1985 \cite{Majthay}, Castrigiano \& Hayes 1993 \cite{C-Hayes}, Golubitsky \& Guillemin  1973 \cite{Gol-G}).  These projections of the catastrophe manifolds are called {\it catastrophe maps} or {\it Lagrangian maps}, and they are differentiably equivalent to  $\bbf_c$ (see \cite[pp. 273-275]{Petters}).  The locally stable
 families $F_{{\bo c},{\bo s}}$ and their induced maps $\bbf_{\bo c}$ are generic for
$n \le 5$, 
and have caustic singularities that are classified according to the parameter $n$.  For $n = 3$, the singularities generically divide into three types: {\it swallowtails}, {\it elliptic umbilics}, and {\it hyperbolic umbilics}.  When $n = 4$, the singularities generically divide into two types: {\it butterflies} and {\it parabolic umbilics}.  
For $n = 5$, they divide into four types: {\it wigwams}, {\it symbolic umbilics}, {\it $2^{nd}$ elliptic umbilics}, and {\it $2^{nd}$ hyperbolic umbilics;} see Table~\ref{table1}.  A detailed treatment of these issues can be found in Poston and Stewart 1978 \cite{Pos-Stewart},
Gilmore 1981 \cite{Gilmore},
\cite{Majthay}, Arnold 1986 \cite{Arnold86}, 
and \cite[Chap. 7]{Petters}.

\mbox{}

\begin{table}
\footnotesize
\centering
\vskip 6pt
\begin{tabular}{| c | c |}
\hline
& \\
& \,\,\,$F_{{\bo {\rm s}}}(x,y) = -s_1 x + s_2 y + \frac{1}{2}x^2 - \frac{1}{3}y^3$\,\,\, \\
Fold (2D) & \\
& \,\,\,\,${\bf f}(x,y)=\left(x\ ,\ y^2\right)$\,\,\,\, \\
& \\
\hline
& \\
& \,\,\,$F_{{\bo {\rm s}}}(x,y) = -s_1 x + s_2 y +\frac{1}{2}x^2 - \frac{1}{2}s_1 y^2 - \frac{1}{4} y^4$\,\,\, \\
Cusp (2D) & \\
& \,\,\,\,${\bf f}(x,y)=\left(x\ ,\ xy+y^3\right)$\,\,\,\, \\
& \\
\hline
& \\
& \,\,\,$F_{c,{\bo {\rm s}}}(x,y) = s_1 x +s_2 y +c(x^2+y^2)+x^3-3xy^2$\,\,\, \\
Elliptic Umbilic (3D) & \\
& \,\,\,\,${\bf f}_{c}(x,y)=\left(3y^2-3x^2-2cx\ ,\ 6xy-2cy\right)$\,\,\,\, \\
& \\
\hline
& \\
& \,\,\,$F_{c,{\bo {\rm s}}}(x,y) = s_1 x +s_2 y +cxy +x^3 + y^3$\,\,\, \\
Hyperbolic Umbilic (3D) & \\
& \,\,\,\,${\bf f}_{c}(x,y)=\left(-3x^2-cy\ ,\ -3y^2-cx\right)$\,\,\,\, \\
& \\
\hline
& \\
& \,\,\,$F_{c,{\bo {\rm s}}}(x,y) = s_1 x - s_2 y -\frac{1}{2}s_2 x^2 + \frac{1}{2}y^2 - \frac{1}{3}c x^3 - \frac{1}{5}x^5$\,\,\, \\
Swallowtail (3D) & \\
& \,\,\,\,${\bf f}_{c}(x,y)=\left(xy + cx^2 + x^4\ ,\ y\right)$\,\,\,\, \\
& \\
\hline
& \\
& \,\,\,$F_{{\bo c},{\bo {\rm s}}}(x,y) = x^6 + c_1 x^4 + c_2 x^3 + s_2 x^2 + s_1 x + \frac{1}{2}y^2-s_2 y$\,\,\, \\
Butterfly (4D) & \\
& \,\,\,\,${\bf f}_{\bo c}(x,y)=\left(-2xy-3c_2 x^2-4c_1 x^3-6x^5\ ,\ y\right)$\,\,\,\, \\
& \\
\hline
& \\
& \,\,\,$F_{{\bo c},{\bo {\rm s}}}(x,y) = x^2 y + y^4 +c_1 x^2 + c_2 y^2 -s_1 x - s_2 y$\,\,\, \\
Parabolic umbilic (4D) & \\
& \,\,\,\,${\bf f}_{{\bo c}}(x,y)=\left(2c_1 x+2xy\ ,\ 2c_2 y+x^2+4y^3\right)$\,\,\,\, \\
& \\
\hline
& \\
& \,\,\,\,\,\,$F_{{\bo c},{\bo {\rm s}}}(x,y) = x^7 + c_1 x^5 + c_2 x^4 + c_3 x^3 + s_2 x^2 + s_1 x + \frac{1}{2}y^2-s_2 y$\,\,\,\,\,\, \\
Wigwam (5D) & \\
& \,\,\,\,${\bf f}_{{\bo c}}(x,y)=\left(-2xy-3c_3 x^2-4c_2 x^3-5c_1 x^4-7x^6\ ,\ y\right)$\,\,\,\, \\
& \\
\hline
& \\
& \,\,\,$F_{{\bo c},{\bo {\rm s}}}(x,y) = x^3+y^4+c_1 x y^2 + c_2 x y + c_3 y^2 +s_2 y + s_1 x$\,\,\, \\
Symbolic umbilic (5D) & \\
& \,\,\,\,${\bf f}_{{\bo c}}(x,y)=\left(-3x^2-c_1 y^2-c_2 y\ ,\ -4y^3-2c_1 x y -c_2 x-2c_3 y \right)$\,\,\,\, \\
& \\
\hline
& \\
& \,\,\,$F_{{\bo c},{\bo {\rm s}}}(x,y) = x^2 y - y^5 + c_1 y^4 + c_2 y^3 + c_3 y^2 + s_2 y + s_1 x$\,\,\, \\
$2^{{\rm nd}}$ Elliptic umbilic (5D) & \\
& \,\,\,\,${\bf f}_{{\bo c}}(x,y)=\left(-2xy\ ,\ -x^2+5y^4-4c_1 y^3-3c_2y^2-2c_3 y\right)$\,\,\,\, \\
& \\
\hline
& \\
& \,\,\,$F_{{\bo c},{\bo {\rm s}}}(x,y) = x^2 y + y^5 + c_1 y^4 + c_2 y^3 + c_3 y^2 + s_2 y + s_1 x$\,\,\, \\
$~~$$2^{{\rm nd}}$ Hyperbolic umbilic (5D) $~~$& \\
& \,\,\,\,${\bf f}_{{\bo c}}(x,y)=\left(-2xy\ ,\ -x^2-5y^4-4c_1 y^3-3c_2y^2-2c_3 y\right)$\,\,\,\, \\
& \\
\hline
\end{tabular}
\caption{For each type of caustic singularity listed, the second column shows the 
universal local forms of the smooth $n$-parameter family of general functions $F_{{\bo c},\bs}$,
along with their $(n-2)$-parameter family of induced general maps $\bbf_{\bo c}$ between planes.  The numbers $2D, 3D$, etc., denote the codimension of the given singularity.}
\label{table1}
\end{table}


\section{Statement of Main Theorem}
\label{Theorem}

For the generic families of functions $F_{{\bo c}, \bs}$ in Table~\ref{table1}, we define the {\it magnification} $\fkM(\bx_i;\bs)$ at a critical point $\bx_i$ of $F_{{\bo c}, \bs}$ relative
to $\bx = (x,y)$ by the reciprocal of the Gaussian curvature at the point
$(\bx_i,F_{{\bo c},\bs}(\bx_i))$ in the graph of $F_{{\bo c},\bs}$:
$$
\fkM({\bx_i; \bs})
= \frac{1}{{\rm Gauss}(\bx_i,F_{{\bo c},\bs}(\bx_i))}\ \cdot
$$
This makes it clear that the magnification invariants established in our theorem are {\it geometric} invariants.  

\vskip 12pt

\begin{theorem}
\label{theorem-main}
For any of the universal, smooth $n$-parameter family of 
general functions $F_{{\bo c},\bs}$
{\rm(}or general mappings $\bbf_{\bo c}${\rm)} in
Table~\ref{table1},
and for
any non-caustic point $\bf s$ (light source position)
in the indicated region, the following results hold for
$\fkM_i \equiv  \fkM({\bx_i; \bs})$:

\begin{enumerate}
\item $A_2$ {\rm (Fold)} Magnification relations in two-image region: 
$$
\fkM_{1} + \fkM_{2} = 0\ .
$$
\item $A_3$ {\rm (Cusp)} Magnification relations in three-image region:
$$
\fkM_{1} + \fkM_{2} + \fkM_3 =0\ .
$$
\item $A_4$ {\rm (Swallowtail)}
 Magnification relation in four-image region: 
$$ 
\fkM_{1} + \fkM_{2} + \fkM_3 + \fkM_4 = 0\ .
$$
\item $D_4^{-}$ {\rm (Elliptic Umbilic)}
Magnification relations in four-image region: 
$$
\fkM_{1} + \fkM_{2} + \fkM_3 + \fkM_4 = 0\ .
$$
\item $D_4^{+}$ {\rm (Hyperbolic Umbilic)}
Magnification relations in four-image region: 
$$
\fkM_{1} + \fkM_{2} + \fkM_3 + \fkM_4 = 0\ .
$$
\item $A_5$ {\rm (Butterfly)}
Magnification relation in five-image region: 
$$ 
\fkM_{1} + \fkM_{2} + \fkM_3 + \fkM_4 + \fkM_5 = 0\ .
$$
\item $D_5$ {\rm (Parabolic Umbilic)}
Magnification relations in five-image region: 
$$
\fkM_{1} + \fkM_{2} + \fkM_3 + \fkM_4 + \fkM_5 = 0\ .
$$
\item $A_6$ {\rm (Wigwam)}
Magnification relations in six-image region: 
$$
\fkM_{1} + \fkM_{2} + \fkM_3 + \fkM_4 + \fkM_5 + \fkM_6 = 0\ .
$$
\item $E_6$ {\rm (Symbolic Umbilic)}
 Magnification relation in six-image region: 
$$ 
\fkM_{1} + \fkM_{2} + \fkM_3 + \fkM_4 + \fkM_5 + \fkM_6 = 0\ .
$$
\item $D^-_6$ {\rm ($2^{\rm nd}$ Elliptic Umbilic)}
 Magnification relations in six-image region: 
$$
\fkM_{1} + \fkM_{2} + \fkM_3 + \fkM_4 + \fkM_5 + \fkM_6 = 0\ .
$$
\item $D^+_6$ {\rm ($2^{\rm nd}$ Hyperbolic Umbilic)}
Magnification relations in six-image region: 
$$
\fkM_{1} + \fkM_{2} + \fkM_3 + \fkM_4 + \fkM_5 + \fkM_6 = 0\ .
$$
\end{enumerate}
\end{theorem}
\noindent We use the A, D, E classification notation of Arnold 1973 \cite{Arnold73} in the theorem.  This notation highlights a deep link between the above singularities and Coxeter-Dynkin diagrams appearing in the theory of simple Lie algebras.  As mentioned in the introduction, the fold and cusp magnification relations are known \cite{Blan-Nar,Sch-Weiss92,Zakharov,Petters}.  The magnification relations for the swallowtail, elliptic umbilic, and hyperbolic umbilic were discovered recently in \cite{Aazami-Petters}. 
\vskip 10 pt

\noindent
{\it Remark.} The results of Theorem~\ref{theorem-main} actually apply even 
when the non-caustic point $\bs$ is not in the maximum number of pre-images region.
However, complex pre-images will appear, which are unphysical in gravitational lensing.

%
\section{A Recursive Relation for Coefficients of Coset Polynomials}
\label{sec:recursion}

In this section, we present some notation and a proposition about polynomials
that will yield the Euler trace formula as a corollary. The notation and the latter
are used in the proof of Theorem~\ref{theorem-main}.  The proposition itself is proved
in the Appendix.

We begin with some notation.  Let $\mathbb{C}[x]$ 
be the ring of polynomials over $\mathbb{C}$ and consider a polynomial
$$
\varphi(x) = a_n x^n + \cdots + a_1 x + a_0 \in \mathbb{C}[x]\ .
$$
Suppose that the $n$  zeros $x_1,\dots,x_n$ pf $\varphi (x)$
are distinct (generically, the roots of a polynomial are distinct) and
let  $\varphi'(x)$ be the derivative of $\varphi(x)$.  
Also, let $R \subset \mathbb{C}(x)$ denote the subring of rational functions that are defined at the roots $x_i$ of $\varphi(x)$:
$$
R = \left\{\frac{p(x)}{q(x)}\ :\ p(x), q(x) \in \mathbb{C}[x]\ {\rm and}\ q(x_i) \neq 0\ 
\mbox{for all roots}\ x_i\ \right\}\ \cdot
$$
Let $(\varphi(x))$ be the ideal in $R$ generated by $\varphi(x)$ and denote the cosets of
the quotient ring $\rideal$ using an overbar.  
Below are two basic results that we prove
in the Appendix (see Claim 2)
for the convenience of the reader:
\begin{itemize}
 \item Members of the same coset in $\rideal$ agree on the roots $x_i$ of $\varphi(x)$, that is,  if
$h_1(x)$ and $h_2(x)$ belong to the same coset, then
$h_1(x_i) = h_2(x_i)$.
\item Every rational function $h(x) \in R$ has in its coset $\oli{h(x)} \in \rideal$ a unique polynomial representative $h_*(x)$ of degree less than $n$.
\end{itemize}

\begin{proposition}
\label{prop:recursive}
Consider any polynomial $\varphi(x) = a_n x^n + \cdots + a_1 x + a_0 
\in \mathbb{C}[x]$
with distinct
roots and any rational function $h(x)\in R$.  Let 
$$
h_*(x) = c_{n-1} x^{n-1} + \cdots + c_1 x + c_0
$$
be the unique polynomial representative of the coset $\oli{h(x)} \in \rideal$ and let
$$
r(x) = b_{n-1} x^{n-1} + \cdots + b_1 x + b_0
$$
be the unique polynomial representative of
the coset $\oli{\varphi'(x)h(x)}\in \rideal$.
Then the coefficients of $r(x)$ are
given in terms of the coefficients of $h_*(x)$ and $\varphi(x)$ 
through the following recursive relation:
\beq
\label{eq:gen-recursive}
b_{n-i} = c_{n-1} b_{n-i,n-1} + \cdots + c_1 b_{n-i,1} + c_0 b_{n-i,0} \,
\hspace{0.75in} i = 1, \dots, n\ , 
\eeq
with
\beq
\label{eq:relations}
\left\{
\begin{array}{ll}
b_{n-i, 0} = (n- (i-1))\, a_{n- (i-1)}\ , & \qquad i = 1, \dots, n\ ,\\
                                       & \\
\displaystyle b_{n-i,k} = -\frac{a_{n-i}}{a_{n}}\, b_{n-1,k-1} + b_{n-(i+1),k-1}\ , & 
                 \qquad  i = 1, \dots, n\ , \qquad  k = 1, \dots, n-1\ ,
\end{array}
\right.
\eeq
where $b_{-1,k-1} \equiv 0$.
\end{proposition}

\noindent By Proposition~\ref{prop:recursive}, if $r_k (x)$ is the unique polynomial representative of the coset $\oli{\varphi'(x)x^k} \in \rideal$, then
\beq\label{eq:rk}
r_k (x) = b_{n-1,k}x^{n-1} + \cdots + b_{1,k} x + b_{0,k}\ , \hspace{1in} k = 0,1, \dots, n-1\ ,
\eeq
where its coefficients are
given in terms of the coefficients of $\varphi(x)$  through (\ref{eq:relations}).

\vskip12pt

\begin{corollary}
\label{Nk1}
Assume the hypotheses and notation of Proposition~\ref{prop:recursive}.
Given the distinct roots $x_1,\dots,x_n$ of $\varphi(x)$, the
Newton sums  
$
N_k \equiv \sum_{i=1}^{n} (x_i)^k
$
satisfy:
\beqa
\label{Nk}
N_k =  \frac{b_{n-1,k}}{a_n}\ ,
\hspace{1in} k = 0,1, \dots, n-1\ .
\eeqa
\end{corollary}
\noindent
In other words, the quantity $a_n \, N_k$ equals the $(n-1)$st coefficient of the unique polynomial representative (\ref{eq:rk}) of the coset $\oli{\varphi'(x)x^k}$ in $\rideal$.

\vskip12pt

\noindent{\bf Proof.} Note that for $k =0$,
eqn. (\ref{eq:relations}) in Proposition~\ref{prop:recursive} yields
$$
b_{n-1,0} = n a_n = N_0 a_n\ .
$$
For $1 \leq k \leq n-1$, there is a known recursive relation for $N_k$, in terms of $N_1,N_2,\dots,N_{k-1}$; see, e.g., Barbeau 1989~\cite[p. 203]{poly}.  It is given by
\beqa
\label{newtonsum}
ka_{n-k} + a_{n-k+1}N_1 + a_{n-k+2}N_2 + \cdots + a_{n-1}N_{k-1} + a_n N_k = 0\ .
\eeqa
We proceed by induction on $k$ for $1 \leq k \leq n-1$. 
For $k=1$, eqn. (\ref{newtonsum}) implies
$
N_1 = - \frac{a_{n-1}}{a_n},
$
while 
eqn. (\ref{eq:relations}) gives
$
b_{n-1,1} = - a_{n-1} = a_n N_1,
$
which agrees with eqn. (\ref{Nk}).  Now assume that
$
b_{n-1,j} = a_n N_j
$
for $j = 1, \dots, k-1.$
To establish the result for $j =k$,
we shall repeatedly apply Proposition~\ref{prop:recursive}:
\beqa
\label{induction}
b_{n-1,k} &=& -\frac{a_{n-1}}{a_{n}}\, b_{n-1,k-1} + b_{n-2,k-1}\nonumber \\
&=& -\frac{a_{n-1}}{a_{n}}\, b_{n-1,k-1} + \left[-\frac{a_{n-2}}{a_{n}}\, b_{n-1,k-2} + b_{n-3,k-2}\right]\nonumber \\
&=& -\frac{a_{n-1}}{a_{n}}\, b_{n-1,k-1} -\frac{a_{n-2}}{a_{n}}\, b_{n-1,k-2} + \left[-\frac{a_{n-3}}{a_{n}}\, b_{n-1,k-3} + b_{n-4,k-3}\right]\nonumber \\
&\vdots&\nonumber \\
&=& -\frac{a_{n-1}}{a_{n}}\, b_{n-1,k-1} -\frac{a_{n-2}}{a_{n}}\, b_{n-1,k-2} -\frac{a_{n-3}}{a_{n}}\, b_{n-1,k-3} - \cdots - \frac{a_{n-(k-1)}}{a_n}\, b_{n-1,1} - \frac{a_{n-k}}{a_n}b_{n-1,0} + b_{n-(k+1),0}\ .\nonumber \\
&=& -\left(a_{n-1}N_{k-1} + a_{n-2}N_{k-2} + a_{n-3}N_{k-3} + \cdots + 
a_{n-(k-1)}N_{1} + ka_{n-k}\right)\nonumber \\
&=& a_nN_k\ , \nonumber
\eeqa
where $b_{n-1,0} = na_n$ and $b_{n-(k+1),0} = (n-k)a_{n-k}$ follow from eqn. (\ref{eq:relations}) in Proposition~\ref{prop:recursive}, and the last equality is due to
(\ref{newtonsum}). $\qed$

\vskip 12pt

\begin{corollary}[Euler Trace Formula]
Assume the hypotheses and notation of Proposition~\ref{prop:recursive}.
For any rational function $h(x) \in R$,  the following holds:
\beqa
\label{euler}
\sum_{i=1}^{n} h(x_i)  = \frac{b_{n-1}}{a_n}\ ,
\eeqa
where 
$b_{n-1}$ is the $(n-1)$st coefficient of the unique polynomial representative 
$r(x)$ of the coset $\oli{\varphi'(x)\, h(x)} \in \rideal$ and
$a_n$ the $n$th coefficient of $\varphi (x)$.

\end{corollary}
\vskip12pt
\noindent {\bf Proof.} 
Let $h_*(x)$ be the unique polynomial
representative of the coset $\oli{h(x)} \in \rideal$.
First note that, since
$h(x)$ and $h_*(x)$ belong to the same coset, we have
$h(x_i) = h_*(x_i)$.  The Euler trace formula now proceeds
from a simple application of Propositon~\ref{prop:recursive} and Corollary~\ref{Nk1}:
\beqa
\sum_{i=1}^{n} h(x_i) &=&  \sum_{i=1}^{n} h_*(x_i)
= \sum_{i=1}^{n} \sum_{j=0}^{n-1} c_j \cdot (x_i)^j = \sum_{j=0}^{n-1} c_j \sum_{i=1}^{n} (x_i)^j
= \sum_{j=0}^{n-1} c_j N_j\nonumber \\
&=& c_{n-1} N_{n-1} \,+\, \cdots\,+\, c_1 N_1 \,+\, c_0 N_0\nonumber \\
&=& c_{n-1} \left(\frac{b_{n-1,n-1}}{a_n}\right) \,+\, \cdots + c_1 \left(\frac{b_{n-1,1}}{a_n}\right) \,+\, c_0 \left(\frac{b_{n-1,0}}{a_n}\right)
\hspace{0.5in} \mbox{(by Corollary~\ref{Nk1})}\nonumber \\
&=& \frac{c_{n-1}b_{n-1,n-1} + \cdots + c_1 b_{n-1,1} + c_0 b_{n-1,0}}{a_n}\nonumber \\
&=& \frac{b_{n-1}}{a_n}\ \cdot    \hspace{0.5in} \mbox{(by Proposition~\ref{prop:recursive})} \nonumber 
\hfill \qed
\eeqa

\vskip12pt

\noindent
{\it Remark.} Dalal \& Rabin 2001 \cite{Dalal-Rabin} gave a different proof of the
Euler trace formula, one employing residues.

\section{Proof of the Main Theorem}
\label{Proof}

We begin by establishing some preliminaries before starting the computational
part of the proof.
Given a family of functions $F_{{\bo c},{\bs}}$,  a parameter vector
$(\bc_0,\bs_0)$ is called a
{\it caustic point} of the family if there is at least one critical point $\bx_0$
of $F_{{\bo c}_0,{\bs_0}}$ (i.e., $\bx_0$ satifies $\grad F_{\bc_0,\bs_0}(\bx_0) = {\bf 0}$)
such that the Gaussian curvature at
$(\bx_0,F_{\bc,\bs}(\bx_0))$ in the graph of $F_{\bc,\bs}$
vanishes.  Furthermore, for the list of singularities in Table~\ref{table1}, the mappings 
$\bbf_{\bf c}$ are induced by the families $F_{{\bo c},{\bs}}$.
In fact, a direct computation shows that for all the singularities,
we can obtain $\bbf_{\bo c}$ through the gradient of
$F_{{\bo c},{\bs}}$ as follows:
$$
\grad F_{\bc,\bs} (\bx) = {\bf 0} \quad \Longleftrightarrow \quad
\bbf_{\bo c} (\bx) = \bs\ .
$$
We can also express the magnification in terms of the general mappings $\bbf_{\bo c}$ induced from the $n$-parameter family of functions $F_{{\bo c},{\bs}}$.  To do so, recall that the Gaussian curvature at a point
$(\bx,F_{\bc,\bs}(\bx))$ in the graph of $F_{\bc,\bs}$ is given
by
$${\rm Gauss}(\bx,F_{{\bo c},\bs}(\bx))
= \frac{\det(\Hess F_{{\bo c},\bs})(\bx)}{1 + |\grad F_{{\bo c},\bs}(\bx)|^2}\ \cdot
$$
At a critical point $\bx_0$, the magnification of $\bx_0$ is then given by
$$
\fkM({\bx_0; \bs})
= \frac{1}{{\rm Gauss}(\bx_0,F_{{\bo c},\bs}(\bx_0))}
= \frac{1}{\det(\Hess F_{{\bo c},\bs})(\bx_0)}\ \cdot
$$
Note that caustics are characterized by the family $F_{{\bo c},\bs}$ having
 at least one infinitely magnified critical point.  A computation also shows that for all the singularities in Table~\ref{table1}, the
following holds:
$$
\det(\Jac \bbf_{\bo c}) = \det(\Hess F_{{\bo c},\bs})\ .
$$
Consequently, we can also express the magnification at a pre-image $\bx_i$ 
of $\bs$ under $\bbf_{\bf c}$ as
$$
\fkM_i \equiv \fkM (\bx_i; \bs)
= \frac{1}{\det(\Jac \bbf_{\bo c})(\bx_i)}\ , \hspace{1in} \bbf_{\bf c} (\bx_i) = \bs\ .
$$
Observe that caustics in the target plane of $\bbf_\bc$
are given equivalently as
points $\bs$ where the Jacobian determinant of $\bbf_\bc$ vanishes.
Now, given an induced mapping $\bbf_{\bo c}$ and a target point
$\bs = (s_1,s_2)$, we can use the pair of equations
$$
(s_1,s_2) = \bbf_{\bo c}(x,y) \equiv (f_{1, {\bf c}}(x,y) , f_{2, {\bf c}}(x,y))
$$
to solve for $(x,y)$ in terms of $(s_1,s_2)$, which will give the pre-images $\bx_i = (x_i,y_i)$
of $\bs$  under $\bbf_{\bo c}$.

For the singularities in
Table~\ref{table1}, we shall see that the pre-images can be determined from solutions
of a polynomial in one variable, which is obtained by eliminating
one of the pre-image coordinates, say $y$.  In doing so we obtain a polynomial $\varphi(x) \in \mathbb{C}[x]$ whose roots will be the $x$-coordinates $x_i$ of the different pre-images 
under $\bbf_{\bo c}$:
$$
\varphi(x) = a_n x^n + \cdots + a_1 x + a_0\ .
$$
Generically, we can assume that the roots of $\varphi (x)$ are
distinct, an assumption made throughout the paper.

We would then be able to express the magnification $\fkM(x,y; \bs)$ at
a general pre-image point $(x,y)$ 
as a function of one variable, in this case $x$, so that
$$
\fkM(x,y(x);\bs) = \frac{1}{J(x,y(x))} \equiv \frac{1}{J(x)} \equiv \fkM(x)\ ,
$$
where 
$
J \equiv \det(\Jac \bbf_{\bo c})
$
and the explicit notational dependence on $\bs$ is dropped for simplicity.
Since we shall consider only non-caustic target points $\bs$
giving rise to pre-images $(x_i, y(x_i))$, we know that $J(x_i) \neq 0$.  
Furthermore, we shall only consider non-caustic points that yield the maximum number
of pre-images. 
In addition, for the singularities in Table~\ref{table1}, the rational function 
$\fkM(x)$ is defined at the roots of $\varphi (x)$, i.e.,
$\fkM(x) \in R$.  
Now, denote by $\fkm(x)$ the unique polynomial representative in the coset 
$\oli{\varphi'(x)\, \fkM(x)} \in \rideal$, and let $b_{n-1}$ be its $(n-1)$st
 coefficient.  
In the notation of Proposition~\ref{prop:recursive}, we have
$h(x) \equiv \fkM(x)$ and 
$r (x) \equiv \fkm(x)$.
Euler's trace formula (Corollary~\ref{euler}) then tells us immediately that the total signed magnification satisfies 
\beq\label{eq:magsum}
\sum_{i} \fkM_i = \frac{b_{n-1}}{a_{n}}\ \cdot
\eeq
It therefore remains to determine the coefficient $b_{n-1}$ for each caustic singularity
in Table~\ref{table1}.  Next to each singularity below we indicate the value of $n-1$,
which is the codimension of the singularity.

Finally, we mention that the full theorem is not a direct consequence of the Euler-Jacobi formula, of multi-dimensional residue integral methods, or of Lefschetz fixed point theory, because some of the singularities have fixed points at infinity.

\begin{enumerate}

\item{\bf Fold ($1$):}  Its corresponding induced map $\bbf$ is
$$
\bbf(x,y) = (x\,,\,y^2)\ .
$$
Let $\bs = (s_1,s_2)$ be a non-caustic point and let us determine the maximum number
of images of $\bs$.  Setting 
$$
\bbf(x,y) = (s_1,s_2)\ ,
$$
we obtain $x = s_1$ and find that the $y$-coordinates of the pre-images are
the two real zeros of the polynomial
$$
\varphi(y) \equiv y^2 - s_2\ .
$$
Consequently, there is a maximum of two pre-images.  
The magnification, expressed in the one variable $y$, is 
given by $\fkM(y) = 1/J(y) = 1/2y$.  Since $\varphi'(y) = 2y = J(y)$, we have
$$
\varphi'(y)\fkM(y) = J(y)\fkM(y) = 1\ .
$$
But this implies that the unique polynomial representative in the coset $\oli{\varphi'(y)\fkM(y)}$ is the polynomial $\fkm(y) \equiv 1$.  Since the $(n-1)$st coefficient is 
$b_1 = 0$, we conclude via (\ref{eq:magsum}) that the total signed magnification in the two-image region is zero:
$$
\fkM_1 + \fkM_2 = \sum_{i=1}^{2} \fkM(y_i) = \frac{b_1}{a_2} = 0\ .
$$

\item{\bf Cusp ($2$):}  Its corresponding induced map $\bbf$ is
$$
\bbf(x,y) = (x\,,\,xy + y^3)\ .
$$
As with the fold, 
let $\bs = (s_1,s_2)$ be a non-caustic point and let us determine the maximum number
of images of $\bs$.
Setting 
$$
\bbf(x,y) = (s_1,s_2)\ ,
$$
we obtain $x = s_1$ and find that the $y$-coordinates of the pre-images are the
three real zeros of the polynomial
$$
\varphi(y) \equiv y^3 + s_1 y - s_2\ .
$$
So, there is a maximum of three pre-images.  
The magnification, expressed in the one variable $y$, is given by $\fkM(y) = 1/J(y) = 1/(3y^2+s_1)$.  Once again we have $\varphi'(y) = J(y)$, so that
$$
\varphi'(y)\fkM(y) = J(y)\fkM(y) = 1\ .
$$
As with the fold, it follows that $\fkm(y) \equiv 1$ and $b_{n-1} = b_2 = 0$, so that the total signed magnification in the three-image region is zero:
$$
\fkM_1 + \fkM_2 + \fkM_3 = \sum_{i=1}^{3} \fkM(y_i) = \frac{b_2}{a_3} = 0\ .
$$

\item{\bf Swallowtail ($3$):}  Its corresponding 1-parameter induced map $\bbf_c$ is
$$
\bbf_{c}(x,y) = (xy + cx^2 + x^4\,,\,y)\ .
$$
Let $\bs = (s_1,s_2)$ be a non-caustic point and let us determine the maximum number
of images of $\bs$.
Setting 
$$
\bbf(x,y) = (s_1,s_2)\ ,
$$
we obtain $y= s_2$ and find that the $x$-coordinates of the pre-images are the
four real zeros of the polynomial
$$
\varphi(x) \equiv x^4 + cx^2 + s_2 x -s_1\ ,
$$
which gives a maximum of four pre-images.  The magnification is $\fkM(x) = 1/J(x) = 1/(4x^3 + 2cx +s_2)$ and $\varphi'(x) = J(x)$, so that
$$
\varphi'(x)\fkM(x) = J(x)\fkM(x) = 1\ .
$$
We thus have $\fkm(x) \equiv 1$, $b_{n-1} = b_3 = 0$, and the total signed magnification in the four-image region is zero:
$$
\fkM_1 + \fkM_2 + \fkM_3 +\fkM_4 = \frac{b_3}{a_4} = 0\ .
$$

\item{\bf Elliptic Umbilic ($3$):} Its corresponding 1-parameter induced map $\bbf_c$ is
$$
{\bbf_{\bo c}}(x,y) = (3 y^2 - 3 x^2 - 2 c x\,,\,6 x y - 2 c y)\ .
$$
Setting 
$
\bbf(x,y) = (s_1,s_2)
$
for a non-caustic point $(s_1,s_2)$, we get
$y = s_2/(6x - 2c)$, which we use to get the following degree $4$ polynomial for
the $x$-coordinates of the pre-images:
$$
\varphi(x) \equiv 4 c^2 s_1 - 3 s_2^2 + (8 c^3 - 24 c s_1) x - (36 c^2 - 36 s_1) x^2 + 108 x^4\ .
$$
Hence there is a maximum of four pre-images. 
The magnification is $\fkM(x,y) = 1/(4 c^2 - 36 x^2 - 36 y^2)$, which becomes a function
of $x$:
$$
 \fkM(x) = \frac{1}{J(x)} = \frac{1}{4 c^2 - 12 s_1 - 24 c x - 72 x^2}\ \cdot
$$
This time $\varphi'(x) \neq J(x)$, but the situation is remedied if we 
multiply through by $2c-6x$ to get
$$
\varphi'(x)\fkM(x) = \left[J(x)\fkM(x)\right](2c-6x) = 2c-6x\ .
$$
Thus the unique polynomial representative in the coset $\oli{\varphi'(x)\fkM(x)}$ is the polynomial $\fkm(x) \equiv -6x+2c$.  Since $b_{n-1} = b_3 = 0$, we have
$$
\fkM_1 + \fkM_2 + \fkM_3 + \fkM_4 = 0\ .
$$

\item{\bf Hyperbolic Umbilic ($3$):} Its corresponding 1-parameter induced map $\bbf_c$, for a given target point $(s_1,s_2)$ lying in the four-image region, is
$$
{\bbf_{\bo c}}(x,y) = (-3 x^2 - c y\,,\,-3 y^2 - c x) = (s_1,s_2)\ .
$$
We eliminate $y$ to obtain a polynomial in the variable $x$, given by
$$
\varphi(x) = -3 s_1^2 - c^2 s_2 - c^3 x - 18 s_1 x^2 - 27 x^4.
$$
The magnification is $\fkM(x,y) = 1/(-c^2 + 36 x y)$.  Substituting for $y$ via ${\bbf_{\bo c}}(x,y) = (s_1,s_2)$, we obtain
$$
 \fkM(x) = \frac{1}{J(x)} = \frac{c}{-c^3 - 36 s_1 x - 108 x^3}\ \cdot
$$
It follows that $cJ(x) = \varphi'(x)$, so that
$$
\varphi'(x)\fkM(x) = \left[J(x)\fkM(x)\right]c = c\ .
$$
Thus the unique polynomial representative in the coset $\oli{\varphi'(x)\fkM(x)}$ is the polynomial $\fkm(x) \equiv c$.  Since $b_{n-1} = b_3 = 0$, we have
$$
\fkM_1 + \fkM_2 + \fkM_3 + \fkM_4 = 0\ .
$$

\item{\bf Butterfly ($4$):} Its corresponding $2$-parameter induced map $\bbf_{\bo c}$, for a given target point $(s_1,s_2)$ lying in the five-image region,  is
$$
{\bbf_{\bo c}}(x,y) = (-2xy-3c_2 x^2-4c_1 x^3-6x^5\,,\,y) = (s_1,s_2)\ .
$$
We eliminate $y$ and are left with a polynomial in the variable $x$,
$$
\varphi(x) \equiv -s_1 -2s_2 x-3c_2 x^2-4c_1 x^3-6x^5\ ,
$$
whose roots are the $x$-coordinates of the five pre-images.  In this case $J(x) = \varphi'(x)$, so that 
$$
\varphi'(x)\fkM(x) = 1\ .
$$
Thus $\fkm(x) \equiv 1$ is the unique polynomial representative in the coset $\oli{\varphi'(x)\fkM(x)}$.  Since $b_{n-1} = b_4 = 0$, it follows that
$$
\fkM_1+\fkM_2+\fkM_3+\fkM_4+\fkM_5 = 0\ .
$$

\item{\bf Parabolic Umbilic ($4$):} Its corresponding 2-parameter induced map $\bbf_{\bo c}$, for a given target point $(s_1,s_2)$ lying in the five-image region, is
$$
{\bbf_{\bo c}}(x,y) = (2c_1x+2xy\,,\,2c_2y+x^2+4y^3) = (s_1,s_2)\ .
$$
We eliminate $x$ to obtain a polynomial in the variable $y$,
\beqa
\label{varphipar}
\varphi(y) \equiv -s_1^2 + 4c_1^2 s_2-(8 c_1^2 c_2 - 8 c_1 s_2)y-(16 c_1 c_2 - 4s_2)y^2-(16 c_1^2+8 c_2)y^3 - 32 c_1 y^4- 16 y^5\ .
\eeqa
The magnification is $\fkM(x,y) = 1/(4 c_1 c_2 - 4 x^2 + 4 c_2 y + 24 c_1 y^2 + 24 y^3)$.  Substituting for $x^2$ via the equations ${\bbf_{\bo c}}(x,y) = (s_1,s_2)$, we obtain
$$
\fkM(y) = \frac{1}{J(y)} = \frac{1}{4 c_1 c_2 - 4s_2 + 12 c_2 y + 24 c_1 y^2 + 40y^3}\ \cdot
$$
Although $\varphi'(y) \neq J(y)$, the situation is remedied if we multiply through by $-2c_1-2y$\,:
\beqa
\label{parabolic2}
\varphi'(y)\fkM(y) = \left[J(y)\fkM(y)\right](-2c_1-2y) = -2c_1-2y\ ,
\eeqa
from which we immediately conclude that $\fkm(y) \equiv -2c_1-2y$ is the unique polynomial representative in the coset $\oli{\varphi'(y)\fkM(y)}$.  Since $b_{n-1}=b_4 = 0$, we have
\beqa
\fkM_1 + \fkM_2 + \fkM_3 + \fkM_4 + \fkM_5 = 0\ .\nonumber
\eeqa

\item{\bf Wigwam ($5$):}  Its corresponding $3$-parameter induced map $\bbf_{\bo c}$, for a given target point $(s_1,s_2)$ lying in the six-image region,  is
$$
{\bbf_{\bo c}}(x,y) = (-2 x y - 3 c_3 x^2 - 4 c_2 x^3 - 5 c_1 x^4 - 7 x^6\,,\,y) = (s_1,s_2)\ .
$$
We eliminate $y$ and are left with a polynomial in the variable $x$,
$$
\varphi(x) \equiv -2 s_2 x  - 3 c_3 x^2 - 4 c_2 x^3 - 5 c_1 x^4 - 7 x^6-s_1\ ,
$$
whose roots are the $x$-coordinates of the six pre-images.  In this case $J(x) = \varphi'(x)$, so that 
$$
\varphi'(x)\fkM(x) = 1\ .
$$
Therefore, as with the fold, cusp, swallowtail, and butterfly, we conclude that $\fkm(x) \equiv 1$ is the unique polynomial representative in the coset $\oli{\varphi'(x)\fkM(x)}$.  Since 
$b_{n-1}=b_5 = 0$, it follows that
$$
\fkM_1+\fkM_2+\fkM_3+\fkM_4+\fkM_5 + \fkM_6= 0\ .
$$

\item{\bf Symbolic Umbilic ($5$):} Its corresponding 3-parameter induced map $\bbf_{\bo c}$, for a given target point $(s_1,s_2)$ lying in the six-image region, is
$$
{\bbf_{\bo c}}(x,y) = (-3 x^2 - c_1 y^2 - c_2 y\,,\,-4 y^3 - 2 c_1 x y - c_2 x - 2 c_3 y) = (s_1,s_2)\ .
$$
We eliminate $x$ to obtain a polynomial in $y$,
\beqa
\varphi(y) &=& -c_2^2 s1 - 3 s_2^2 - c_2^3 y - 4 c_1 c_2 s_1 y - 12 c_3 s_2 y - 
 5 c_1 c_2^2 y^2 - 12 c_3^2 y^2 - 4 c_1^2 s_1 y^2\nonumber \\
 &-& 8 c_1^2 c_2 y^3- 24 s_2 y^3 - 4 c_1^3 y^4 - 48 c_3 y^4 - 48 y^6\ . \nonumber
 \eeqa
The magnification is $\fkM(x,y) = 1/(-c_2^2 + 12 c_3 x + 12 c_1 x^2 - 4 c_1 c_2 y - 4 c_1^2 y^2 + 72 x y^2)$.  Substituting for $x$ via the equations $\bbf_{\bo c}(x,y) = (s_1,s_2)$ gives
$$
\fkM(y) = \frac{1}{J(y)} = \frac{c_2 + 2 c_1 y}{-c_2^3 - 4 c_1 c_2 s_2 - 12 c_3 s_2 - (10 c_1 c_2^2 + 24 c_3^2 + 8 c_1^2 s_2) y - (24 c_1^2 c_2 + 72 s_2) y^2 - (16 c_1^3 + 192 c_3) y^3 - 288 y^5}\ \cdot
$$
It is not difficult to check that $\varphi'(y) = J(y)(2c_1 y+c_2)$, so that
$$
\varphi'(y)\fkM(y) = \left[J(y)\fkM(y)\right](2c_1 y+c_2) = 2c_1 y+c_2\ .
$$
Then $\fkm(y) \equiv 2c_1 y + c_2$ and $b_{n-1}=b_5 = 0$, so that
$$
\fkM_1 + \fkM_2 + \fkM_3 + \fkM_4 + \fkM_5 + \fkM_6 = 0\ .
$$

\item{\bf ${\bo 2^{nd}}$ Elliptic Umbilic ($5$):} Its corresponding 3-parameter induced map $\bbf_{\bo c}$, for a given target point $(s_1,s_2)$ lying in the six-image region, is
$$
{\bbf_{\bo c}}(x,y) = (-2xy\,,\,-x^2 + 5 y^4 - 4 c_1 y^3 - 3 c_2 y^2 - 2 c_3 y) = (s_1,s_2)\ .
$$
Eliminating $x$, we obtain the polynomial
$$
\varphi(y) \equiv -s_1^2 - 4 s_2 y^2 - 8 c_3 y^3 - 12 c_2 y^4 - 16 c_1 y^5 + 20 y^6\ .
$$
The magnification is $\fkM(x,y) = 1/(-4 x^2 + 4 c_3 y + 12 c_2 y^2 + 24 c_1 y^3 - 40 y^4)$.  Substituting for $x^2$ via the equations $\bbf_{\bo c}(x,y) = (s_1,s_2)$ gives
$$
\fkM(y) = \frac{1}{J(y)} = \frac{1}{4 s_2 + 12 c_3 y + 24 c_2 y^2 + 40 c_1 y^3 - 60 y^4}\ \cdot
$$
One can check directly that $\varphi'(y) = J(y)(-2y)$, so that
$$
\varphi'(y)\fkM(y) = \left[J(y)\fkM(y)\right](-2y) = -2y\ .
$$
Then $\fkm(y) \equiv -2y$ and $b_{n-1}=b_5 = 0$, so that
$$
\fkM_1 + \fkM_2 + \fkM_3 + \fkM_4 + \fkM_5 + \fkM_6 = 0\ .
$$

\item{\bf ${\bo 2^{nd}}$ Hyperbolic Umbilic ($5$):} Its corresponding 3-parameter induced map $\bbf_{\bo c}$, for a given target point $(s_1,s_2)$ lying in the six-image region, is
$$
{\bbf_{\bo c}}(x,y) = (-2xy\,,\,-x^2 - 5 y^4 - 4 c_1 y^3 - 3 c_2 y^2 - 2 c_3 y) = (s_1,s_2)\ .
$$
Eliminating $x$, we obtain the polynomial
$$
\varphi(y) \equiv -s_1^2 - 4 s_2 y^2 - 8 c_3 y^3 - 12 c_2 y^4 - 16 c_1 y^5 - 20 y^6\ .
$$
The magnification is $\fkM(x,y) = 1/(-4 x^2 + 4 c_3 y + 12 c_2 y^2 + 24 c_1 y^3 + 40 y^4)$.  Substituting for $x^2$ via the equations $\bbf_{\bo c}(x,y) = (s_1,s_2)$ gives
$$
\fkM(y) = \frac{1}{J(y)} = \frac{1}{4 s_2 + 12 c_3 y + 24 c_2 y^2 + 40 c_1 y^3 + 60 y^4}\ \cdot
$$
Once again, it is easy to check that $\varphi'(y) = J(y)(-2y)$, so that
$$
\varphi'(y)\fkM(y) = \left[J(y)\fkM(y)\right](-2y) = -2y\ .
$$
Then $\fkm(y) \equiv -2y$ and $b_{n-1}= b_5 = 0$, so that
$$
\fkM_1 + \fkM_2 + \fkM_3 + \fkM_4 + \fkM_5 + \fkM_6 = 0\ .
$$
\end{enumerate}
This completes the proof. $\qed$

\section{Conclusion}
\label{Conclusion}
The paper presented a theorem about the magnification pre-images for caustic
singularities up to codimension five.  We proved that 
for generic
families of general mappings between planes locally exhibiting such singularities, and for any point in the target lying in the region giving rise to the maximum number of real pre-images, the total signed magnification of the pre-images sums to zero.  The signed magnifications
are invariants as they are Gaussian curvatures at critical points. 
Our result extends earlier work that considered the case of singularities 
through to codimension three. 
The  proof of the theorem is algebraic in nature
and utilizes the Euler trace formula.  In fact, we established
a proposition that relates the coefficients of the unique polynomial in
the coset of certain rational funtions to Newton sums. 
It was then shown that the Euler trace formula follows readily as a 
corollary of our proposition.
The findings of the paper are expected to be relevant to the study of dark matter substructures
on galactic scales using gravitational lensing.  In addition, since the results 
hold for generic general mappings, they are applicable 
to any system in which stable caustic singularities appear.

\section{Acknowledgments}
\noindent ABA would like to thank William L. Pardon and Alberto M. Teguia  for helpful discussions.  AOP acknowledges the support of NSF Grant DMS-0707003.

\appendix

\section{Proof of Proposition~\ref{prop:recursive}}

%
%

For convenience, we restate the result:
\vskip 12pt

\noindent
{\bf Proposition.}
Consider any polynomial $\varphi(x) = a_n x^n + \cdots + a_1 x + a_0 
\in \mathbb{C}[x]$
with distinct
roots and any rational function $h(x)\in R$.  Let 
$$
r(x) = b_{n-1} x^{n-1} + \cdots + b_1 x + b_0
$$
be the respective unique polynomial representative of the coset
$\oli{\varphi'(x)h(x)}$ in $\rideal$.
Then the coefficients of $r(x)$ are
given in terms of the coefficients of $h_*(x)$ and $\varphi(x)$ 
through the following recursive relation:
\beq
\label{eq:gen-recursive}
b_{n-i} = c_{n-1} b_{n-i,n-1} + \cdots + c_1 b_{n-i,1} + c_0 b_{n-i,0} \,
\hspace{0.75in} i = 1, \dots, n\ ,
\eeq
with
\beq
\label{relations}
\left\{
\begin{array}{ll}
b_{n-i, 0} = (n- (i-1))\, a_{n- (i-1)}\ , & \qquad i = 1, \dots, n\ ,\\
                                       & \\
\displaystyle b_{n-i,k} = -\frac{a_{n-i}}{a_{n}}\, b_{n-1,k-1} + b_{n-(i+1),k-1}\ , & 
                 \qquad  i = 1, \dots, n\ , \qquad  k = 1, \dots, n-1\ ,
\end{array}
\right.
\eeq
where $b_{-1,k-1} \equiv 0$.
\vskip 12pt

\noindent
{\it Proof of Proposition.} 

\noindent
We begin with some preliminaries about quotient rings to make
the proof more self-contained.
Let $\mathbb{C}[x]$ be
the ring of polynomials over $\mathbb{C}$ and let $\mathbb{C}(x)$ be the field of
rational functions formed from quotients of polynomials in  $\mathbb{C}[x]$.
The $n$ zeros $x_1,\dots,x_n$ of
$\varphi(x) = a_n x^n + \cdots + a_1 x + a_0 \in \mathbb{C}[x]$
are assumed to be distinct (generically, the roots of a polynomial are distinct). 
Let $(\varphi(x))$ denote the ideal in $\mathbb{C}[x]$ generated by $\varphi(x)$, and consider the quotient ring $\cideal$, whose cosets we denote by $\oli{g(x)}$. 
This quotient ring has two important properties:

\begin{itemize}
\item {\it Property 1:} If $\oli{g_{1}(x)} =\oli{g_{2}(x)}$, then by definition $g_{1}(x) - g_{2}(x) = h(x)\varphi(x)$ for some $h(x) \in \mathbb{C}[x]$, from which it follows that $g_{1}(x_{i}) = g_{2}(x_{i})$ for all $n$ roots $x_i$ of $\varphi(x)$.  Thus members of the same coset must agree on the roots of $\varphi(x)$, so that, in particular, $\sum_{i=1}^{n} g_1(x_i) = \sum_{i=1}^{n} g_2(x_i)$.
\item {\it Property 2:} Each coset $\oli{g(x)}$ has a {\it unique} representative of degree 
at most $n-1$, as follows:  by the division algorithm in $\mathbb{C}[x]$, there exist polynomials $q(x)$ and $r(x)$ such that
$$
g(x) = q(x)\varphi(x) + r(x)\ ,
$$
where deg $r < {\rm deg}\ \varphi = n$.  Passing to the quotient ring $\cideal$, we see that $\oli{g(x)} = \oli{r(x)}$.  Suppose now that there exists another polynomial $p(x)$ of degree less than $n$ with $\oli{g(x)} = \oli{p(x)}$.  Then $\oli{p(x)} = \oli{r(x)}$, so that
$$
p(x) - r(x) = h(x)\varphi(x)
$$
for some $h(x) \in \mathbb{C}[x]$.  If $h(x) \not\equiv 0$, then deg $h\,\varphi \geq n$, while the degree of the left-hand side is less than $n$.  We must therefore have $h(x) \equiv 0$ and $p(x) = r(x)$.  We may thus represent every coset by its unique polynomial representative of degree less than $n$, which in turn implies that $\cideal$ is a vector space of dimension $n$, with basis $\left\{ \oli{1},\oli{x},\oli{x^2}\dots,\oli{x^{n-1}}\right\}$.
\end{itemize}

The next result will be used to show that Properties 1 and 2 also hold for a
certain subset of rational functions in $\mathbb{C}(x)$ (see Claim 2 below).

\vskip 12pt

\noindent
{\it Claim 1.}
  Let $x_1,\dots,x_n \in \mathbb{C}$ be distinct.  Let $c_1,\dots,c_n \in \mathbb{C}$, not necessarily distinct.  Then there exists a unique polynomial $H(x) \in \mathbb{C}[x]$ with deg $h < n$ such that $H(x_i) = c_i$.
\vskip 12pt

\noindent
{\it Proof (Claim 1).}
Induction on $n$.  For $n=1$, define $H(x) \equiv c_1$.  Now assume that the result is true for $n-1$, and consider a set of $n$ distinct complex numbers $x_1,\dots,x_n$.  By the induction hypothesis, there exists a polynomial $h(x) \in \mathbb{C}[x]$ with deg $h < n-1$ such that $h(x_i)=c_i$ for $i = 1,\dots,n-1$.  Now define
$$
H(x) = h(x) + \frac{(x-x_1)(x-x_2)\cdots(x-x_{n-1})}{(x_n-x_1)(x_n-x_2)\cdots(x_n-x_{n-1})}\,\left(c_n-h(x_n)\right)\ .
$$
It follows that $H(x) \in \mathbb{C}[x]$ has degree less than $n$, and $H(x_i) = c_i$ for all $i = 1,\dots,n$.  (As a simple example to show that $H(x)$ need not be unique if the $x_1,\dots,x_n$ are not distinct, consider the numbers $2,2,3,3$ all being mapped to $0$.  Then the polynomials $H_1(x) = (x-2)^2(x-3)$, $H_2(x)=(x-2)(x-3)^2$, and $H_3(x) = (x-2)(x-3)$ all satisfy the assumptions of the lemma.)  Suppose that there exist two polynomials $H_1(x)$ and $H_2(x)$ with $H_1(x_i) = c_i = H_2(x_i)$.  By the division algorithm in $\mathbb{C}[x]$, there are unique polynomials $q(x)$ and $r(x)$ such that
$$
H_1(x) - H_2(x) = q(x)\left[(x-x_1)(x-x_2)\cdots(x-x_n)\right] + r(x)\ ,
$$
where deg $r < n$.  If $q(x) \not\equiv 0$, then the degree of the polynomial on the right-hand side is at least $n$, whereas $H_1(x) - H_2(x)$ has degree less than $n$.  We must therefore have $q(x) \equiv 0$.  Moreover, if $r(x) \not\equiv 0$, then $H_1(x_i) = H_2(x_i)$ gives that $r(x_i) = 0$ for all $x_1,\dots,x_n$.  This implies, however, that $r(x)$ has $n$ distinct zeros and so must have degree $n$, a contradiction.  Thus $H_1(x) = H_2(x)$.  $\qed$ (Claim 1)
\vskip 12pt

Let $R \subset \mathbb{C}(x)$ denote the subring of rational functions that are 
defined at the roots $x_i$ of $\varphi(x)$,
$$
R = \left\{\frac{p(x)}{q(x)}\ :\ p(x), q(x) \in \mathbb{C}[x]\ {\rm and}\ q(x_i) \neq 0\ 
\mbox{for all roots}\ x_i\ \right\}\ ,
$$
and consider the quotient ring $\rideal$.  The next claim states that
the ring $\rideal$ satisfies Properties 1 and 2.
\vskip 12pt

\noindent
{\it Claim 2.}
Members of the same coset in $\rideal$ agree on the roots $x_i$ of $\varphi(x)$, that is,  if
$g_1(x)$ and $g_2(x)$ belong to the same coset, then
$g_1(x_i) = g_2(x_i)$, and so
 $\sum_{i=1}^{n} g_1(x_i) = \sum_{i=1}^{n} g_2(x_i)$.
In addition, any rational function $h(x) \in R$ will have in its coset $\oli{h(x)} \in \rideal$ a unique polynomial representative $r(x)$ of degree less than $n$.

\vskip 12pt

\noindent
{\it Proof (Claim 2).}
Notice that, if $\oli{h_1(x)} = \oli{h_2(x)} \in \rideal$, then by definition there exists a rational function $h(x) \in R$ such that
$$
h_1(x) - h_2(x) = h(x)\varphi(x)\ ,
$$
so that $h_1(x_i) = h_2(x_i)$ for all the zeros $x_i$ of $\varphi(x)$.  In other words, $\rideal$ also satisfies Property $1$.  It turns out that when the zeros $x_1,\dots,x_n$ of $\varphi(x)$ are distinct, as we are assuming they are, then $\rideal$ also satisfies Property 2 (in fact $\rideal$ and $\cideal$ will be isomorphic as rings).  For given a coset $\oli{h(x)} \in \rideal$, Claim 1
shows that there is a unique polynomial $g(x) \in \mathbb{C}[x]$  of degree less than $n$ whose values at the $n$ roots $x_i$ are $h(x_i)$.  Then the rational function $g(x)-h(x) \in R$ vanishes at every $x_i$, and a simple application of the division algorithm applied to the numerator of  $g(x)-h(x)$ shows that $\oli{g(x)} = \oli{h(x)} \in \rideal$.  Thus any rational function $h(x) \in R$ will have in its coset $\oli{h(x)} \in \rideal$ a unique polynomial representative $r(x)$ of degree less than $n$. $\qed$ (Claim 2)
\vskip 12pt

We now begin the proof of the Proposition by establishing the following Lemma:
\vskip 12pt

\noindent
{\bf Lemma.}
Let $\varphi(x) = a_n x^n + \cdots + a_1 x + a_0$ and consider the
quotient ring $\rideal$.  For any $1 \leq k \leq n-1$, let 
$$
r_k (x) \equiv b_{n-1,k}\, x^{n-1} + \cdots + b_{1,k}\, x + b_{0,k} 
$$
be the unique polynomial representative in the coset $\oli{\varphi'(x)x^k}$.  Then the following recursive relation holds: 
\beq
\label{relations}
\left\{
\begin{array}{ll}
b_{n-i, 0} = (n- (i-1))\, a_{n- (i-1)}\ , & \qquad i = 1, \dots, n\ ,\\
                                       & \\
\displaystyle b_{n-i,k} = -\frac{a_{n-i}}{a_{n}}\, b_{n-1,k-1} + b_{n-(i+1),k-1}\ , & 
                 \qquad  i = 1, \dots, n\ , \qquad  k = 1, \dots, n-1\ ,
\end{array}
\right.
\eeq
where $b_{-1,k-1} \equiv 0$.
\vskip 12pt

\noindent
{\it Proof of Lemma.}
The existence and uniqueness of the polynomial
$$
r_k (x) = b_{n-1,k}\,{x^{n-1}} + \cdots + b_{n-i,k} \, {x^{n-i}} + \cdots + b_{1,k} \, {x} + b_{0,k}\, {1}\ ,
$$
where
\beqa
\label{eq:bi-a}
\oli{\varphi'(x)\, x^k} = \oli{r_k(x)} = 
b_{n-1,k}\, \oli{x^{n-1}} + \cdots + b_{n-i,k}\, \oli{x^{n-i}} + \cdots + b_{1,k} \oli{x} + b_{0,k}\, \oli{1}\ ,
\eeqa
were established in Claim 2.  Also, note that since 
$\oli{\varphi(x)} = \oli{0} \in \rideal$, we have
\beqa\label{eq:xn-coset-a}
\oli{x^{n}} = -\frac{a_{n-1}}{a_n}\oli{x^{n-1}} - \cdots - \frac{a_1}{a_n}\oli{x} - \frac{a_0}{a_n}\oli{1}\ .
\eeqa

\noindent
\underline{Case $k = 0$:} By (\ref{eq:bi-a}),
we get
$$
\oli{\varphi'(x)x^0} =  \oli{r_0(x)}
= b_{n-1,0}\, \oli{x^{n-1}} 
+ \cdots + b_{n-i,0}\, \oli{x^{n-i}} + \cdots
+ b_{1,0}\, \oli{x} + b_{0,0}\, \oli{1}\ .
$$
However,
$$
 \oli{\varphi'(x)x^0}
= \oli{\varphi'(x)}
= na_n\, \oli{x^{n-1}} + \cdots + (n-(i-1))a_{n-(i-1)}\, \oli{x^{n-i}} + \cdots + 2 a_2\, \oli{x} 
+ a_1\,  \oli{1}\ .
$$
Consequently,
\beqa
\label{eq:b0-b1-a}
b_{n-i, 0} = (n- (i-1))\, a_{n- (i-1)}\ , \qquad i = 1, \dots, n\ .
\eeqa

\noindent
\underline{Case $k = 1, \dots, n-1$:} 
Equations (\ref{eq:bi-a}) and (\ref{eq:xn-coset-a}) yield
\beqa
\oli{\varphi'(x)\, x^k} &=& b_{n-1,k}\, \oli{x^{n-1}} + \cdots + b_{n-i,k} \, \oli{x^{n-i}} + \cdots + b_{1,k}\, \oli{x} + b_{0,k}\, \oli{1} \nonumber \\
&=& \oli{x \, \varphi'(x)\, x^{k-1}} \nonumber \\
&=& \oli{x} \left[b_{n-1,k-1}\, \oli{x^{n-1}} + b_{n-2,k-1}\, \oli{x^{n-2}} + \cdots + b_{1,k-1} \, \oli{x} + b_{0,k-1}\, \oli{1} \right] \nonumber \\
&=& b_{n-1,k-1}\, \oli{x^{n}} + b_{n-2,k-1}\, \oli{x^{n-1}} + \cdots + b_{1,k-1}\, \oli{x^2} + b_{0,k-1}\, \oli{x} \nonumber \\
&=& b_{n-1,k-1}\left[-\frac{a_{n-1}}{a_n}\, \oli{x^{n-1}} - \cdots - \frac{a_1}{a_n}\, \oli{x} - \frac{a_0}{a_n}\, \oli{1}\right] + b_{n-2,k-1}\, \oli{x^{n-1}} + 
\cdots + b_{1,k-1}\, \oli{x^2} + b_{0,k-1}\, \oli{x} \nonumber \\
&=& \sum_{i=1}^{n} \left[-\frac{a_{n-i}}{a_n}\, b_{n-1,k-1} + b_{n-(i+1),k-1}\right]\, \oli{x^{n-i}}\ . \nonumber 
\eeqa
The coefficients of (\ref{eq:bi-a}) are then related to the coefficients of $a_i$ of $\varphi(x)$
as follows:
$$
b_{n-i,k} = -\frac{a_{n-i}}{a_{n}}\, b_{n-1,k-1} + b_{n-(i+1),k-1}\ , \qquad 
i = 1, \dots, n\ , \qquad 
k = 1, \dots, n-1\ ,
$$
where the coeffiencients $b_{n-i,0}$ are given by
(\ref{eq:b0-b1-a}). Note that $b_{n,k} = 0$ since the unique
polynomial goes up to degree $n-1$. $\qed$ (Lemma)

We now complete the proof of the Proposition.
If $h_{1,*}(x)$ and $h_{2,*}(x)$ are the unique polynomial representatives of the cosets $\oli{h_1(x)}$ and $\oli{h_2(x)}$, respectively, then by uniqueness,
 the sum $h_{1,*}(x) + h_{2,*}(x)$ is the unique polynomial representative of the coset $\oli{h_1(x) + h_2(x)}$.  With that said, we note that, since $\oli{h(x)} = \oli{h_*(x)}$, it follows that $\oli{r(x)} = \oli{\varphi'(x)h(x)} = \oli{\varphi'(x)h_*(x)}$.  We thus have
\beqa
\label{r:coeff}
\oli{r(x)} &=& \oli{\varphi'(x)h_{*}(x)}\nonumber \\
&=& \oli{c_{n-1}\varphi'(x)x^{n-1}} + \cdots + \oli{c_1\varphi'(x)x} + \oli{c_0\varphi'(x)}\nonumber \\
&=& \oli{c_{n-1}r_{n-1}(x)} + \cdots + \oli{c_1r_1(x)} + \oli{c_0r_0(x)}\nonumber \\
&=& c_{n-1} \sum_{i=1}^{n} b_{n-i,n-1}\oli{x^{n-i}} + \cdots + c_{1} \sum_{i=1}^{n} b_{n-i,1}\oli{x^{n-i}} + c_{0} \sum_{i=1}^{n} b_{n-i,0}\oli{x^{n-i}}\nonumber \\
&=& \sum_{i=1}^n \left(c_{n-1}b_{n-i,n-1} + \cdots + c_1b_{n-i,1} + c_0b_{n-i,0}\right)\,\oli{x^{n-i}}\nonumber \\
&=& \sum_{i=1}^n b_{n-i}\oli{x^{n-i}}\ .~~~\qed \ \mbox{(Proposition)} \nonumber
\eeqa


\end{document}